\documentclass[epj]{webofc}
\usepackage[varg]{txfonts}   
\usepackage{subfigure}
\wocname{EPJ Web of Conferences}
\woctitle{CONF12}
\begin{document}

\selectlanguage{english}
\title{The Muon g-2 experiment at Fermilab}
%
%

\author{Antoine Chapelain\inst{1}\fnsep\thanks{\email{antoine.chapelain@cornell.edu} }
On behalf of the Muon g-2 Collaboration}

\institute{Cornell University, Department of Physics, 511 Clark Hall, Ithaca, NY 14853, USA}

\abstract{The upcoming Fermilab E989 experiment will measure the muon anomalous magnetic \mbox{moment $a_{\mu}$.} This measurement is motivated by the 
	  previous measurement performed in 2001 by the BNL E821 experiment that reported a 3-4 standard deviation discrepancy between 
          the measured value and the Standard Model prediction. The new measurement at Fermilab aims to improve the precision by a factor of four
	  reducing the total uncertainty from 540 parts per billion (BNL E821) to 140 parts per billion (Fermilab E989). 
	  This paper gives the status of the experiment.}
\maketitle

\section{Introduction}

The measurement of the anomalous part $a_{\ell=e,\mu}$ of the magnetic moment of the electron and muon is a high-precision test of the Standard Model (SM) of particle 
physics, and thus a good window to look for physics that lies beyond the SM (BSM). The sensitivity to BSM physics scales with $(m_\ell / \Lambda)^2$, $\Lambda$ being 
the BSM physic energy scale,. The current best measurement of $a_e$~\cite{a_e} to 0.28 parts per 
trillion is 1929 more precise than the current best measurement of $a_\mu$~\cite{bnl} to 0.54 parts per million (ppm). 
The muon to electron mass ratio of \mbox{$(m_\mu/m_e)=200$} results an expected BSM sensitivity ratio \mbox{$(m_\mu/m_e)^2=40000$}. 
In consequence, the muon sector is a more sensitive probe to look for BSM physic.


The measurement of $a_\mu$  reported in~\cite{bnl} by the 
BNL E821 experiment shows a 3.6 standard deviation difference from the SM prediction, which could be a sign for BSM physics such as 
supersymmetry~\cite{Fargnoli, Bach} or  dark photon~\cite{Batley}. This tantalizing difference motivated the new experiment E989 at Fermilab by the Muon g-2 
collaboration \cite{E989}, which aims for a factor of four improvement in the total experimental uncertainty. 
E821 total systematic uncertainty was dominated by the statistical uncertainty. E989 aims to have the same contribution to the total uncertainty from
systematics and statistics. It requires a much larger improvement for the statistical uncertainty with 21 times the E821 statistics.
The total uncertainty improvement is 
discussed in Sections~\ref{sec:stat},~\ref{sec:omega} and \ref{sec:field}. Section~\ref{sec:th} below gives an overview of the theory status and 
Section~\ref{sec:exp} presents the measurement principle.

\section{Theory}
\label{sec:th}

A charged fermion with electric charge $Q= \pm 1$ under the influence of a magnetic field $B$ has a magnetic dipole moment 
\begin{equation} \vec{\mu}=g \frac{Q}{2m} \vec{s},\end{equation}
where $\vec{s}$ is the spin of the particle 
and $g$  the gyromagnetic ratio. 
We define the anomalous magnetic moment via the deviation of the gyromagnetic ratio from Eq. (1): $g = 2(1+a_\mu)$. 
In the Dirac theory, which corresponds to leading order in QED
, $g=2$, and $a_\mu = 0$. Higher order contributions 
lead to $a_\mu \neq 0$. The current theoretical status for $a_\mu$ is presented in Table~\ref{tab:sm_pred} \cite{E989}. 
The main contribution comes by far from QED, which is known to five  loops (tenth order) and has a small, well-understood uncertainty.
Sensitivity at the level of the electroweak (EW) contribution was reached by the E821 experiment.
The hadronic contribution dominates the uncertainty (0.43 ppm compared to 0.01 ppm for QED and EW grouped together). 
This contribution splits into two categories, hadronic vacuum polarization (HVP) and hadronic light-by-light (HLbL).  The 
HVP contribution dominates the correction, and can be calculated from $e^+e^- \rightarrow $ hadrons cross-section using dispersion relations.  
The HLbL contribution derives from model-dependent calculations.
Lattice QCD predictions of these two hadronic contributions are becoming competitive, and will be crucial in providing robust uncertainty 
estimates free from uncontrolled modeling assumptions.
Lattice QCD predictions have well-understood, quantifiable uncertainty estimates.  
Model-based estimates lack controlled uncertainty estimates, and will always allow a loophole in comparisons with the SM.

\begin{table}[t]
\begin{center}
{\small
\begin{tabular}{lr}
\hline \hline {\sc\small } &
{\sc\small  Value $(\times \, 10^{-11})$ units  }
\\ \hline
QED & $116\,584\,718.95\pm 0.08$ \\
HVP & $6\,850.6 \pm 43$\\
HLbL&  $105\pm 26 $ \\
EW & $153.6\pm 1.0 $ \\
 \hline
 Total SM & $116\,591\,828 \pm 49$\\
 \hline\hline\
\end{tabular}}
\caption{Summary of the Standard-Model contributions to the muon anomalous magnetic moment~\cite{Blum:2013xva}.}
\label{tab:sm_pred}
\end{center}
\end{table}

The uncertainties in the theory calculation are expected to improve by a factor of two on the timescale of the E989 experiment. 
This improvement will be achieved taking advantage of 
new data to improve both the HVP (BESIII~\cite{BESIII}, VEPP2000~\cite{vepp} and B-factories data) and HLBL (KLOE-2~\cite{kloe} and BESIII data), 
the latter gaining from the modeling improvements made 
possible with the new data. On the lattice QCD side, new ways of computing $a_\mu$ from first principles and an increase in computing capability will 
provide the expected gains. Given the anticipated improvements in both experimental and theoretical precision, if the central values remain the same there is a 
potential $~7$ standard deviation between theory and measurement (5 standard deviation with only experimental improvement). 
The situation is summarized in Fig.~\ref{fig:measpred}~\cite{Blum:2013xva}.

\section{Measurement principle}
\label{sec:exp}

\begin{figure}[!t]
 \subfigure[]{\includegraphics[width=0.5\textwidth]{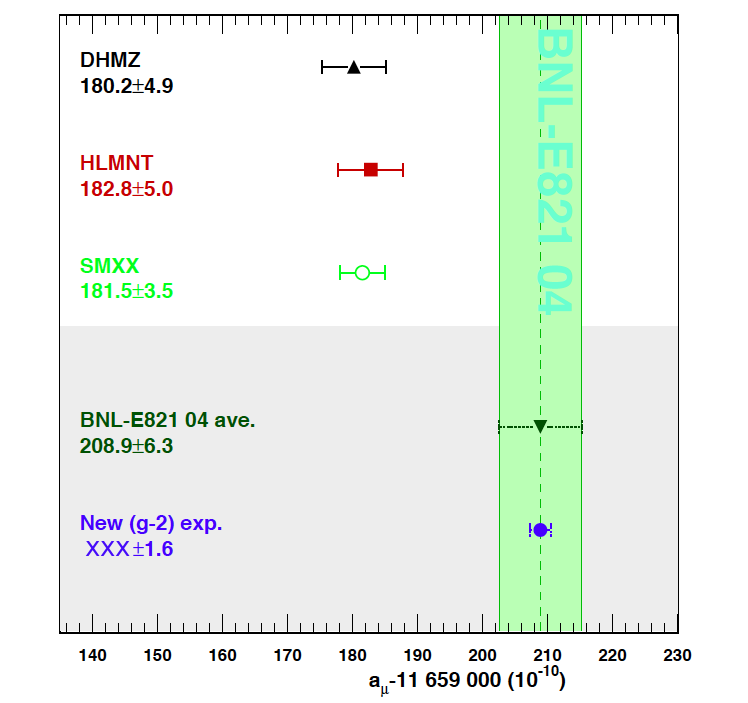}\label{fig:measpred}}  
 \subfigure[]{\raisebox{0.24\height}{\includegraphics[width=0.5\textwidth]{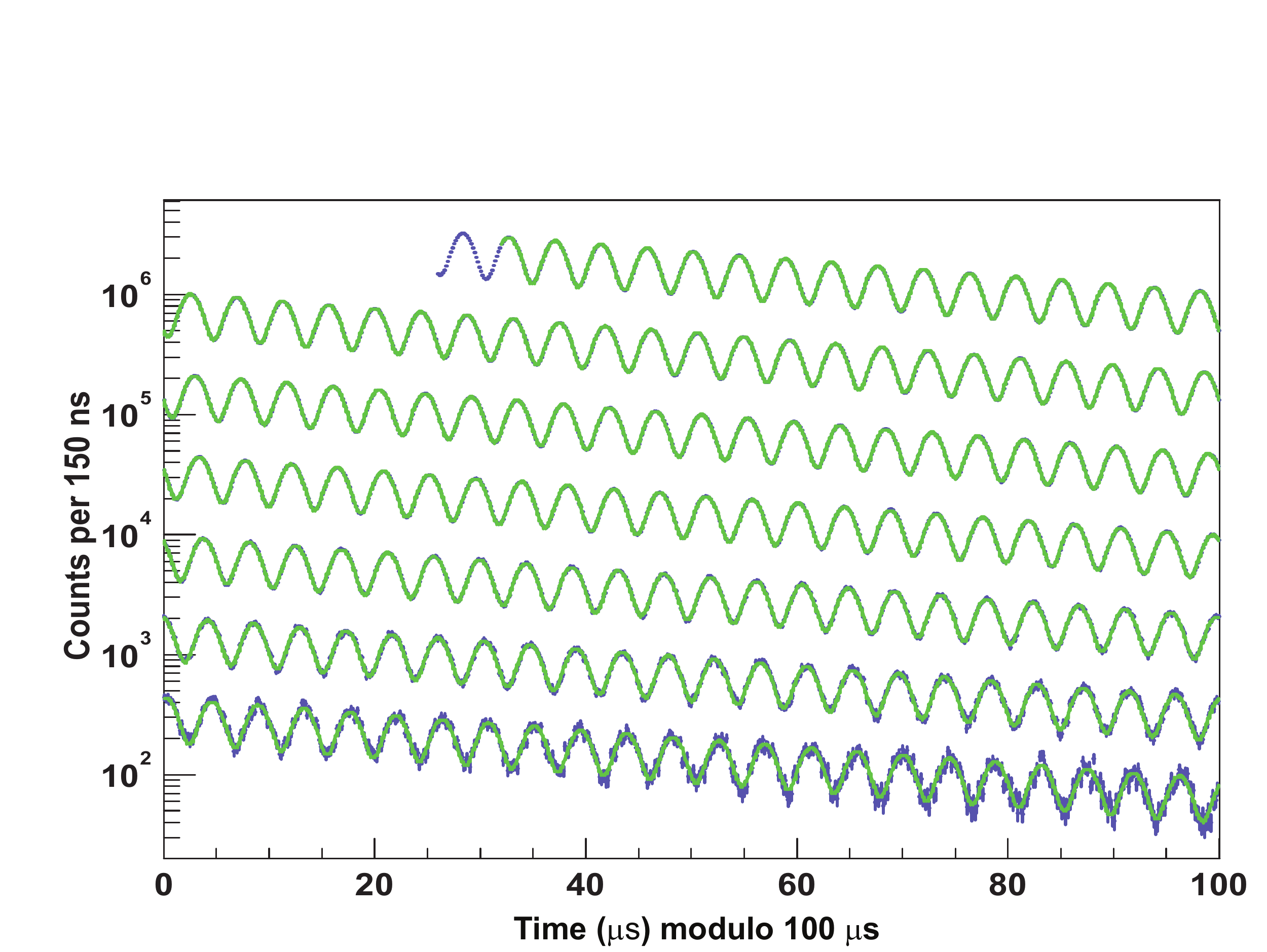}}\label{fig:wiggle}}
\caption{\ref{fig:measpred} Summary of the current SM prediction along with the BNL E821 measurement and the projected uncertainty on the incoming 
Fermilab E989 experiment. \ref{fig:wiggle} Number of high-energy positrons versus time for the 2001 data set from the BNL E821 experiment.}
\label{fig:measurement}
\end{figure}

The E989 experiment at Fermilab follows the same measurement principle that was used by the previous E821 experiment. 
E989 reuses E821 the magnet which is a critical part of the experiment that produces the storage magnetic field to the require very high uniformity. 
A beam of positive muons with 96\% polarization is produced from a beam of pions. The muons are injected into a weak 
focusing storage going through an inflector that cancels the dipole magnetic field of the storage ring. The trajectory of the muons 
at the exit of the inflector does not match the closed orbit of the ring. Three identical kickers will kick the beam onto the closed orbit providing 
a kick of about 11 mrad  corresponding to a magnetic  field of about 230 Gauss. 
The storage magnet only provides the radial focusing. Four electrostatic quadrupoles operating at 32 kV will provide the 
vertical focusing.

Due to the magnetic field, the muon spin precesses relative to its momentum while the muon undergo a cyclotron motion with a frequency of 149 ns. 
Our measurement technique directly yields the anomalous spin precession frequency, the difference between the spin precession frequency and 
the cyclotron frequency, and depends directly on $a_{\mu}$ and the magnetic field strength $B$ via

\begin{equation}
\vec{\omega_a} = \vec{\omega_S} - \vec{\omega_C} = \vec{\omega}_{a}=
 - \frac {Q}  {m} 
\left[  
a_{\mu} \vec{B} 
- \left( a_{\mu} -   \frac {1} {\gamma^2 - 1}   \right)
 \frac {\vec{\beta} \times \vec{E}} {c} \right]
\end{equation}


The magnetic field arising from the boost of any laboratory electric fields to the muon rest frame contributes to the second term
proportional to $\vec{\beta} \times \vec{E}$ in brackets of Eq.~(2).  In E989, 
the focusing quadrupole fields would make such a contribution.  Fortunately, this contribution vanishes at first order for muons having the so-called "magic momentum" of 3.094 GeV/c ($\gamma=29.3$), the momentum of the E821 and E989 
stored beams.

The muon spin precession frequency is extracted using the positron from the muon decay as a proxy. Due to V-A nature of the muon decay, the highest 
energy positrons are emitted 
with their momentum parallel to the muon spin. When this dependence is couple to the boost from the muon rest frame to the lab frame, the rate of 
highest energy positrons in the lab frame experiences a modulation at the anomalous precession frequency $\omega_a$. 
The $\omega_a$ frequency can thus be extracted by counting the number of positrons above an energy threshold (optimized for the best 
sensitivity) as a function of time. Figure~\ref{fig:wiggle} shows the modulated counting rate observed in the 2001 E821 data.
The anomalous magnetic moment is determined measuring both $\omega_a$ and $B$ via the relation
\begin{equation}
 a_\mu  = \frac{\omega_a}{\omega_p}\frac{\mu_p}{\mu_e}\frac{m_{\mu}}{m_e}\frac{g_e}{2},
\end{equation}
where the proton-to-electron magnetic moment ratio ${\mu_p}/{\mu_e}$, the muon-to-electron mass ratio ${m_{\mu}}/{m_e}$ and
the electron gyromagnetic ratio $g_e$ come from other measurements as well as SM theory~\cite{Mohr, Hanneke}.

The projected four-fold improvement on $a_\mu$ requires reducing the statistical uncertainty (limiting uncertainty for E821) by a factor four and the 
systematic uncertainty by a factor three. The following sections focus on these improvements.

\section{First challenge: statistical uncertainty}
\label{sec:stat}

\begin{figure}[!t]
  \centering
  \includegraphics[width=0.5\textwidth]{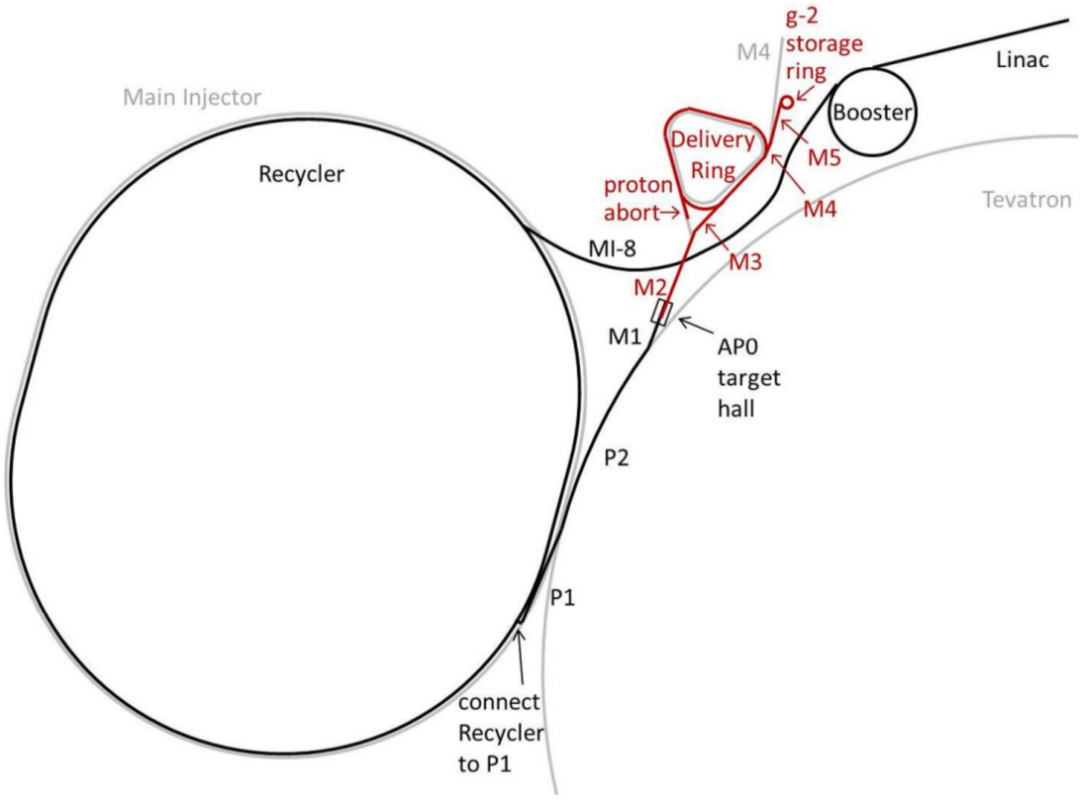}
\caption{Schematic of the Fermilab accelerator chain: from the proton to the muon beam.}
\label{fig:acc}
\end{figure}

The goal for the statistical uncertainty requires 21 times the E821 statistics, which is about \mbox{$2\times10^{11}$} detected positrons. 
This goal represents a great challenge given the 
Fermilab proton beam is 4 times less intense than the BNL proton beam. In consequence, a factor 84 improvement is required for the integrated luminosity. 
To achieve this goal, the muon injection efficiency into the storage will be higher 
for E989 compared to E821. There will be in consequence more muons stored into the ring. Also, the repetition rate  is being increased from 0.37 Hz 
to 12 Hz on average.

Most of the Fermilab 
accelerator chain will be used to produce the muon beam (see Fig.~\ref{fig:acc}). About $4 \times 10^{12}$ 8 GeV protons from the Booster synchrotron will be injected into the Recycler ring 
at a \mbox{15 Hz}  rate. The protons will be re-organized in four bunches with 10 ms time separation and then sent to the pion production target. Pions with a 3.11 GeV/c 
momentum will be 
collected after the target using a Lithium lens and a dipole magnet. The pions are then sent to the Delivery ring. The Delivery ring allows spatial separation 
between the pions (most of them having decayed into muons) and the remaining protons to get rid off the protons. After the Delivery ring, the muon beam goes to the g-2 
storage ring for a typical fill length of 700 $\mu$s. The typical muon injection efficiency is $2.4\times10^{-7}$ per proton on target~\cite{vlad}. 

The accelerator chain described above has a longer $\pi \rightarrow \mu$ decay channel than the one at BNL. E989 will not suffer from a "pion flash" at the early time of the fill that E821 had to deal 
with. It results in a higher muon beam  purity allowing the use of data at an earlier time.  

\section{Second challenge: $\omega_a$ systematic uncertainty}
\label{sec:omega}

\begin{table}[t]
\begin{center}
\begin{tabular}{|l|r|l|c|} \hline
Category & E821  & E989 Improvement Plans & Goal \\
            & [ppb] &              & [ppb] \\ \hline
Intra-fill gain change & 120 & Better laser calibration  & \\ & & Low-energy threshold & 20  \\
Pileup       & 80 & Low-energy samples recorded   & \\ & & Calorimeter segmentation & 40 \\
Lost muons   & 90 & Better collimation in ring  & 20  \\
CBO          & 70 &  Change CBO frequency &  $<30$  \\
$E$ and pitch  & 50 & Improved tracker  & \\ & & Precise storage ring simulations & 30 \\
\hline
Total        & 180 & Quadrature sum & 70 \\
\hline
\end{tabular}
\begin{tabular} {l}
\hline
\end{tabular}
\label{tab:syst}
\caption{List of the major $\omega_a$ systematic uncertainties.}
\end{center}
\end{table}

The main sources of systematic uncertainty $\delta\omega_a$ on $\omega_a$ are listed in Table~\ref{tab:syst} along with their values for E821 
and the design goal for E989. New detector systems 
were designed in order to reduce the total systematic uncertainty on $\omega_a$ by a factor about 3. 

Twenty four calorimeter stations will be deployed inside the ring to measure the energy and arrival time of the positrons. 
The segmented calorimeters (E821's calorimeter was not segmented) are made of an array of $6\times9$ lead-fluoride crystals and provide a fast response due to \u{C}erenkov light. 
Each crystal is readout by a
a 4x4 array of silicon photomultiplier (SiPM), whose signals are summed. The resulting waveform for the crystal is digitized at 800 MAP with 12-bit resolution
(E821 had 200 MSPS 8-bit resolution digitizers). The digitizers are implemented as 5-channel AMC cards that operate in an industry standard ${\mu}$TCA crate.  Etc.


The anticipated main contribution (40 ppb) to $\delta\omega_a$ comes from pile-up 
events when two positrons hit the same 
calorimeter nano-seconds or less apart. The calorimeter segmentation allows spatial separation of pile-up events. The digitization rate allows 
for time separation of pile-up events at nano-seconds time scale.  The energy scale (gain) of the calorimeter will be measured using a laser calibration system. It is important to know accurately the intra-fill 
gain stability since the measurement of $\omega_a$ relies on counting the number of positrons above an energy threshold on a fill basis.
The main design goals for the calorimeter system are:

\begin{enumerate}
\item Positron hit time measurement with an accuracy of 100 psec for a positron energy $>$ \mbox{100 MeV};
\item The measured deposited energy must have a resolution better than 5\% at 2 GeV;
\item Allow 100\% pile-up separation above 5 nsec, and 66 \% below 5 nsec;
\item Gain stability monitored at subpermill level over the course of 700 $\mu$s fill where rate varies by $10^4$. 
\end{enumerate}

\begin{figure}[!t]
  \centering
 \subfigure[]{\includegraphics[width=0.4\textwidth]{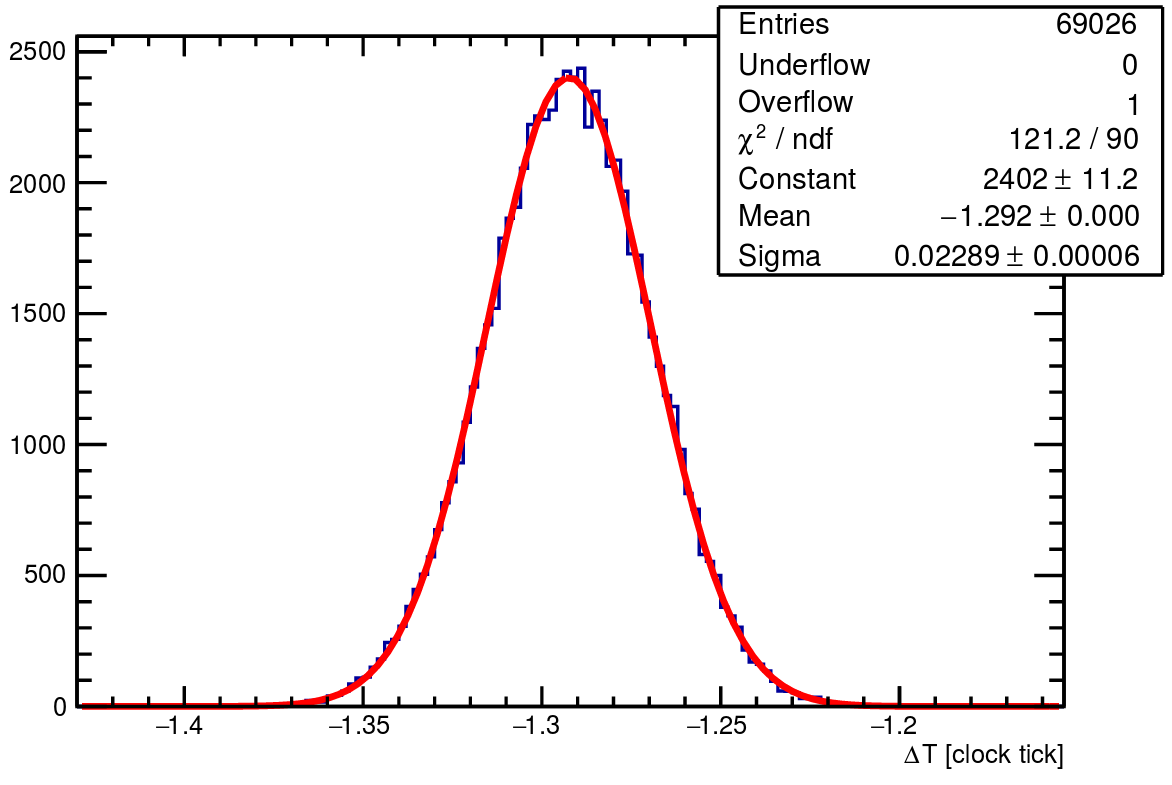}\label{fig:dt}}  
 \subfigure[]{\raisebox{0.1\height}{\includegraphics[width=0.4\textwidth]{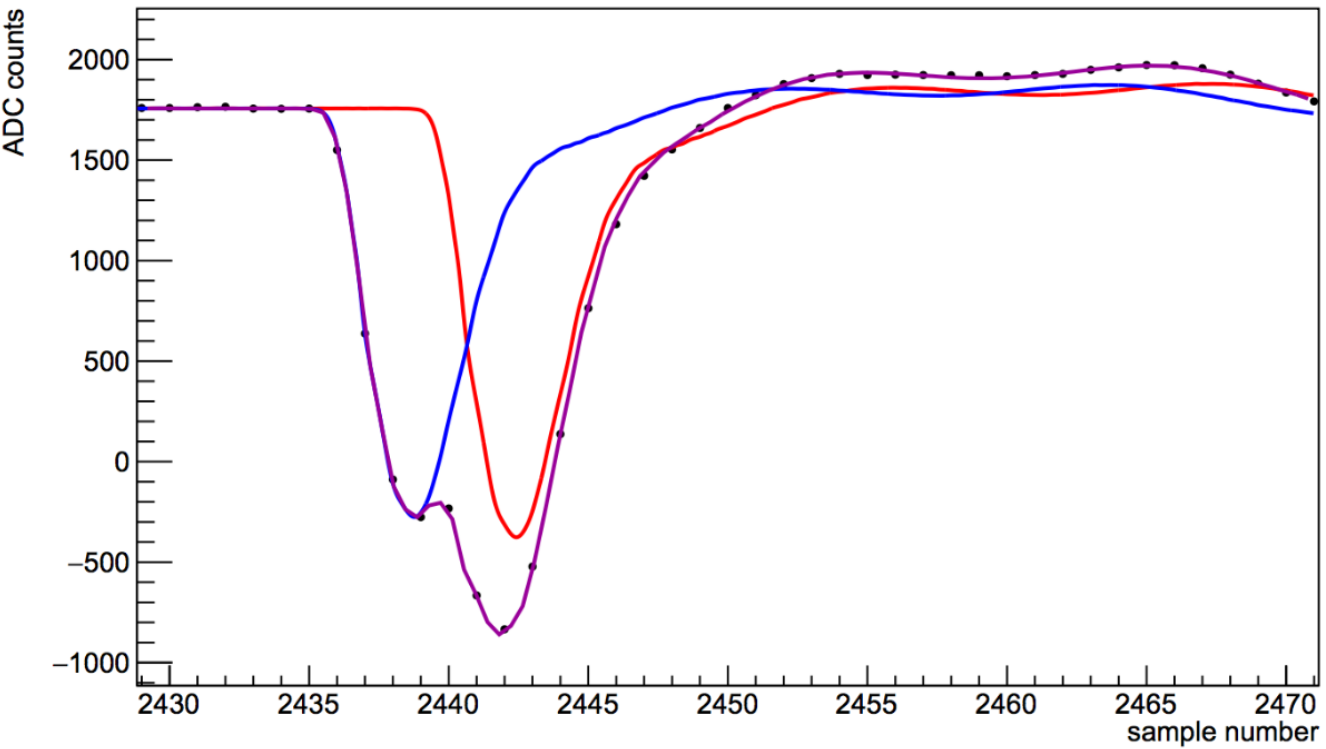}}\label{fig:pileup}}
 \subfigure[]{\includegraphics[width=0.4\textwidth]{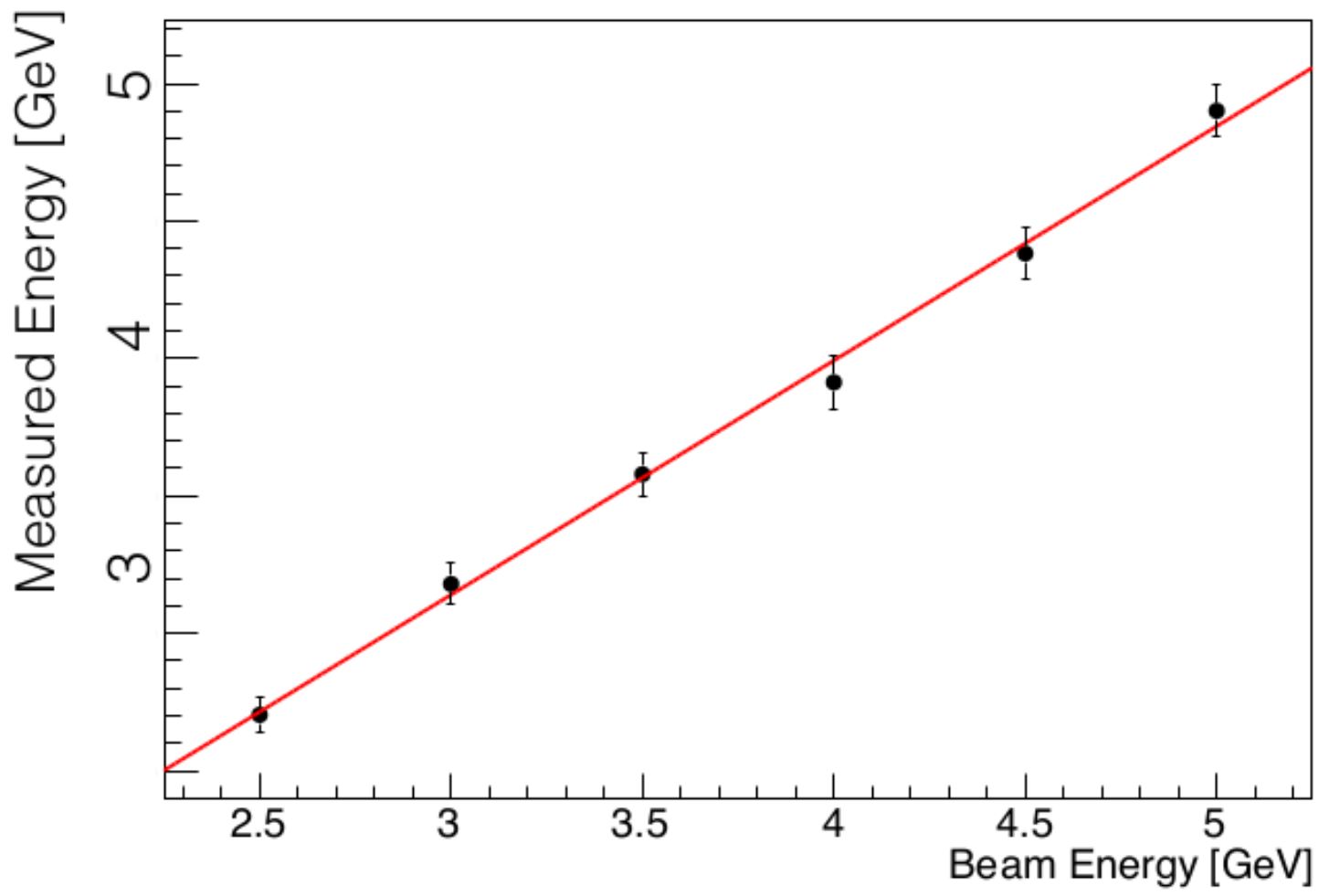}\label{fig:beamE}}
 \subfigure[]{\raisebox{0.25\height}{\includegraphics[width=0.4\textwidth]{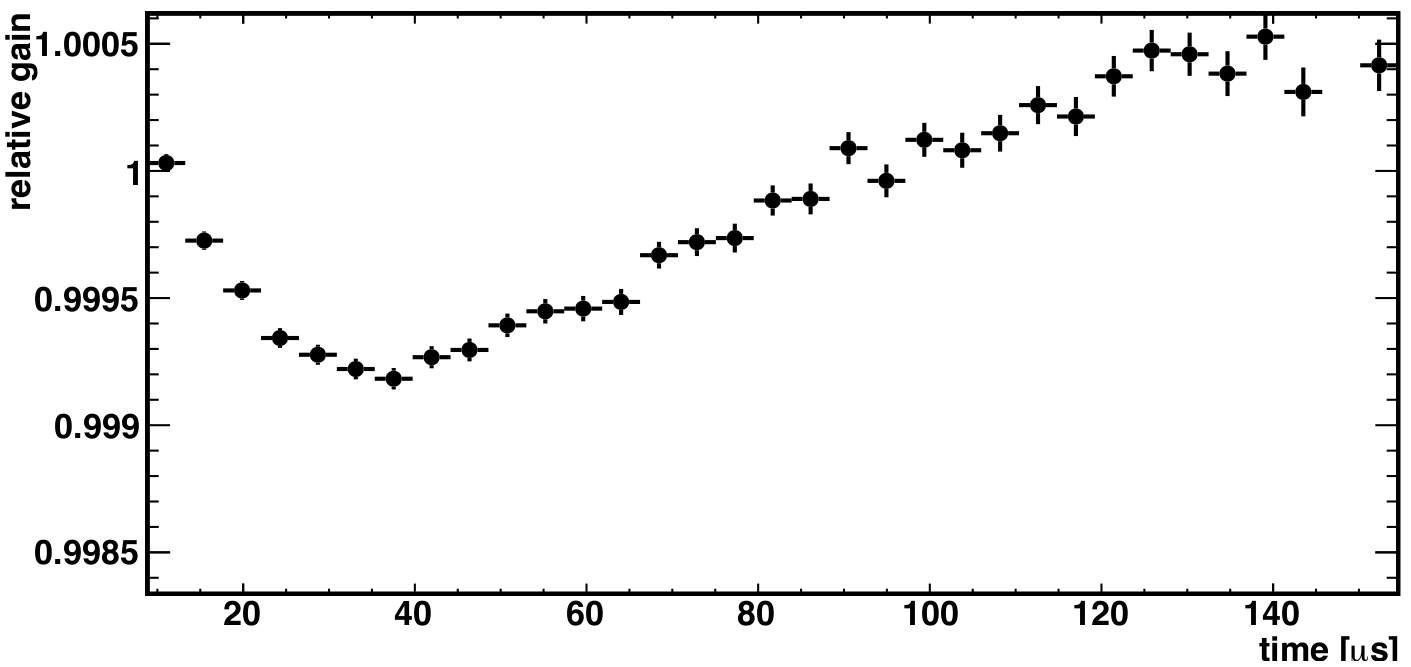}}\label{fig:gain}}
\caption{\ref{fig:dt} Laser light hit time difference between two calorimeter crystal in unit of clock tick. The width of the distribution is the time resolution
considering two crystals at the same time (factor $1/\sqrt2$ gives the resolution for one crystal).
\ref{fig:pileup} Template fitting for a two electron pile-up event with 4.5 ns time interval.
\ref{fig:beamE} Reconstructed beam energy as a function of the true beam energy. The uncertainty on the reconstructed energy is about 3\%.
\ref{fig:gain} Energy scale stability of the calorimeter as a function of time.}
\label{fig:measurement}
\end{figure}

An end-to-end test of one complete calorimeter station was performed in June 2016 with the electron beam at the End Test Station B at the Stanford
Linear Accelerator laboratory (SLAC). 
The electron beam had a primary energy of 3 GeV (ranged from 2.5 GeV to 5 GeV). The number of electron per beam pulse was chosen to be either one, 
two, three... according to the needs for pile-up studies. This test beam showed than the 
design requirements are fulfilled as illustrated in Fig.~\ref{fig:dt}-\ref{fig:beamE}. 
The electron hit time resolution (Fig.~~\ref{fig:dt}) was determined using laser pulses that illuminated all 54 calorimeter crystals simultaneously.  
By comparing the time differences for pulses in different crystals, we found a typical hit time resolution of 25 ps at \mbox{3 GeV}.
Figure~\ref{fig:beamE} shows the linearity of the calorimeter in the GeV range. Each entry for the reconstructed energy has a resolution of about 3\%.
Figure~\ref{fig:pileup} shows a pile-up event with 4.5 ns time separation to which a template fit is applied.  The template fit is able to resolve such 
events and to  reconstruct accurately the energy and hit time of each electron. Finally, Fig.~\ref{fig:gain} shows the evolution of the system gain 
as a function of the rate in the detector, which tell us about the intra-fill gain stability systematic. The laser calibration system was used 
to mimic a realistic fill structure (64 $\mu$s life-time decay in intensity with a factor $10^4$ decay factor over the 700 $\mu$s fill). The laser 
firing rate was a 100 times the expected rate for the experimental running condition at Fermilab. Under these extreme conditions the gain variation meets the goal with $|\delta G|<10^{-3}$.

The data acquisition system (DAQ) for E989 is based on the Maximum Integrated Data Acquisition System (MIDAS) software~\cite{midas}. The DAQ 
must read out data without dead time from more than 1,300 read-out 
channels with 12 Hz fills on average 100 Hz at peak). This represent abouts 20 GB/s of data. The on-line 
processing for the calorimeter data (very large fraction of all the data) uses GPUs processing to only select data of interest, that is, the pulses originated from 
a particle interacting with the calorimeter crystals. This result for the calorimeter data to reduce the rate from 18 GB/s to \mbox{75 MB/s}. Data monitoring is also 
required to ensure on-line and off-line the data quality / integrity. The total anticipated data stored to tape for the experiment is about 2 PB.

The last major system improvement is the tracker system, which will give access to the dynamic of the stored muon beam. There are three tracker stations,
each station being made of 8-module 1500-channel straw tracker. Each station is located in front of one calorimeter. The tracker system will allow control 
of several systematic uncertainties. Within the time window used to count the number of positrons above an energy threshold, some of the stored muons are being lost outside of the 
storage region due to beam dynamic effect. The energy-time correlation of these lost muons changes the polarization content of the counted positrons over time and thus
is a source of systematic uncertainty. The tracker stations will allow monitoring of the lost muons.
The tracker system can also address coherent betatron oscillation (CBO). The tune of the 
closed orbit differs from one. Due to the betatron oscillation and the geometrical acceptance, each calorimeter station will see a different part 
of the beam turn-by-turn. This affect the counting of the positrons. From the measured trajectory of the decay positron, the tracker system can reconstruct the dynamics 
of the beam and its CBO motion. An other important systematic has to do with the vertical focusing of the beam. 
Due to the momentum spread of the beam from the magic momentum, the electrostatic quadrupoles contributes to $\omega_a$ at 
a non-negligible level and need to be corrected for. The tracker system allows to reconstruct the momentum spread of the stored muon beam (or 
equivalent the frequency or radius distribution) that enters the estimation of the correction. There is also a contribution to $\omega_a$ due to the fact that the 
muon have a small but non-zero vertical component to their velocity.

On a different topic, the tracker system makes possible the measurement of the electric dipole moment (EDM) of the muon by looking at the vertical component of the 
momentum of the decay positrons. An EDM would introduce a vertical tilt to the $\omega_a$ vector resulting in an vertical upper/lower asymmetry for the number of 
measured positrons. A factor 2 of improvement in the precision is expected with respect to the E821 measurement.

\section{Third challenge: $\omega_p$ systematic uncertainty}
\label{sec:field}

\begin{figure}[!t]
  \centering
 \subfigure[]{\raisebox{0.23\height}{\includegraphics[width=0.45\textwidth]{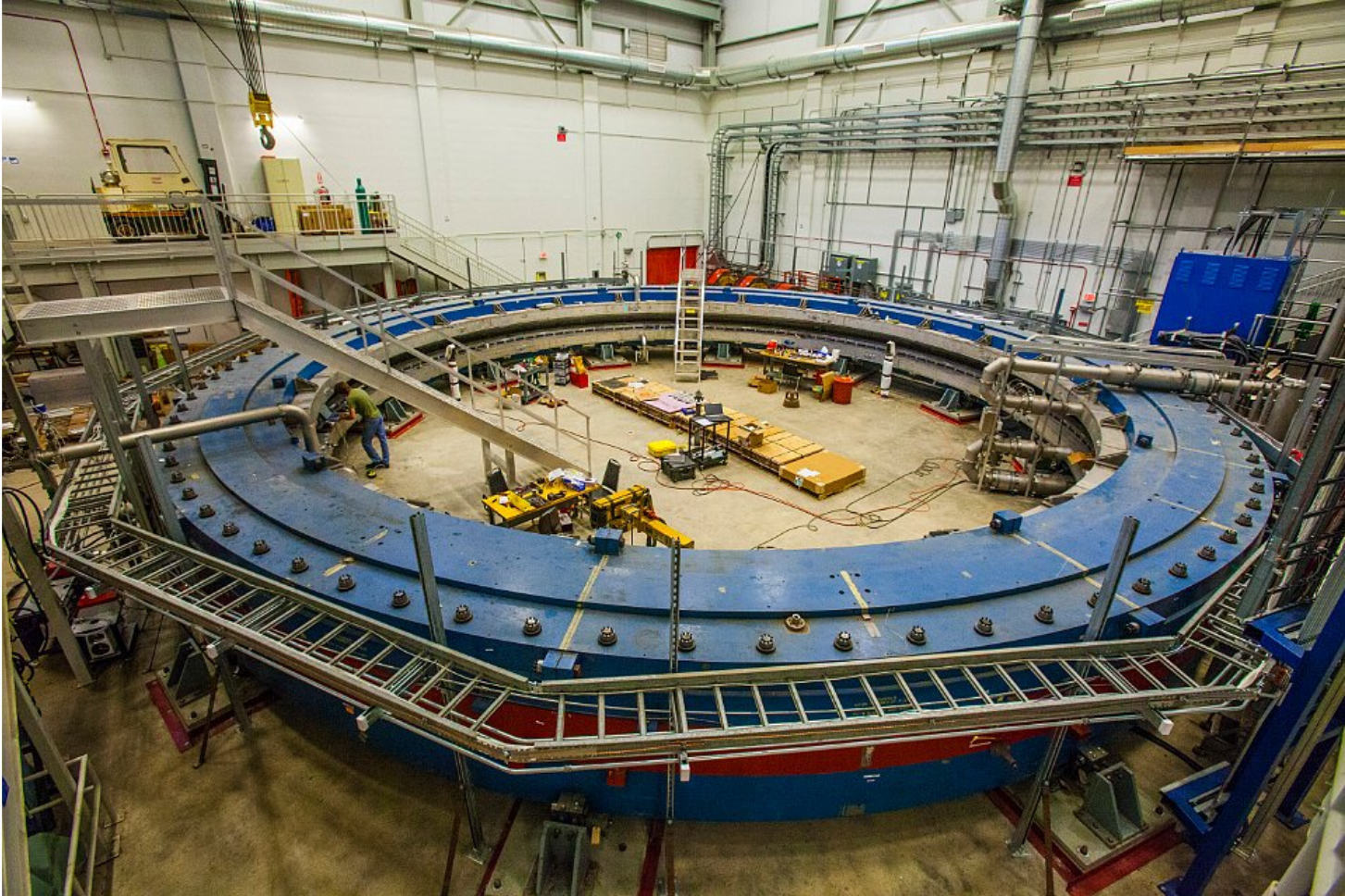}}\label{fig:ring}}  
 \subfigure[]{\raisebox{0.\height}{\includegraphics[width=0.45\textwidth]{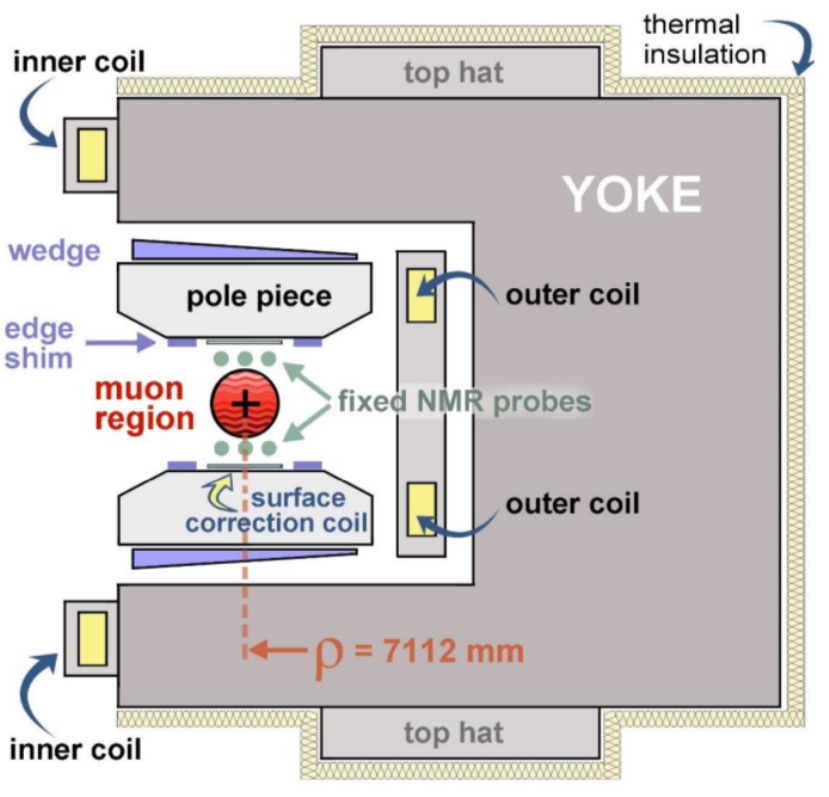}}\label{fig:magnet}}
\caption{\ref{fig:ring} Storage ring in the experimental hall at Fermilab. \ref{fig:magnet} Sketch of the cross-section of the storage magnet.}
\end{figure}

The measurement of the intensity of the magnetic field is the second key measurement to determine the value of $a_\mu$. It is measured in term of the Larmor 
precession frequency of a free proton:
\begin{equation}
\hbar \omega_{p} = 2 \mu_{p} | \overrightarrow{B}|.
\end{equation}
The magnet used by E821 is reused for the current experiment. It is a continuous 44.7 m circumference magnet made of 12 30-degree C-shaped iron yokes and 
superconducting coils (see Fig.~\ref{fig:ring},~\ref{fig:magnet}). 5200 Amp circulate through the coils to produce a 1.45 T field. Other important pieces (poles, wedges, top hats...) are part 
of the magnet to help improving the field uniformity as discussed below. E989 aims to know the magnetic field averaged over running time and the muon 
distribution to an uncertainty of $\pm$70 ppb. It requires:

\begin{enumerate}
\item Producing as uniform magnetic field as possible by shimming the magnet;
\item Stabilizing B in time at the sub-ppm level by feedback, with mechanical and thermal stability;
\item Monitoring B to the 20 ppb level around the storage ring during data collection;
\item Periodically mapping the field throughout the storage region and correlating the field
map to the monitoring information without turning off the magnet between data collection and field mapping. It is essential that the magnet not be powered off unless
absolutely necessary;
\item Obtaining an absolute calibration of the B-field relative to the Larmor frequency of
the free proton.
\end{enumerate}

\begin{figure}[t]
  \centering
 \subfigure[]{{\includegraphics[width=0.9\textwidth]{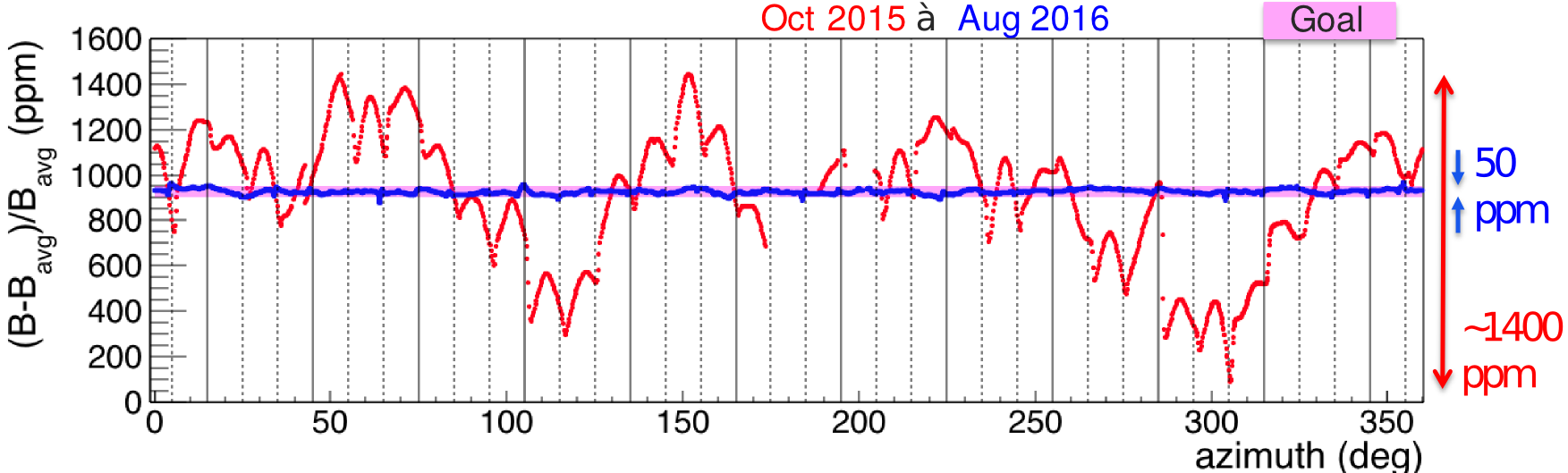}}\label{fig:b1}}  
 \subfigure[]{\raisebox{0.\height}{\includegraphics[width=0.45\textwidth]{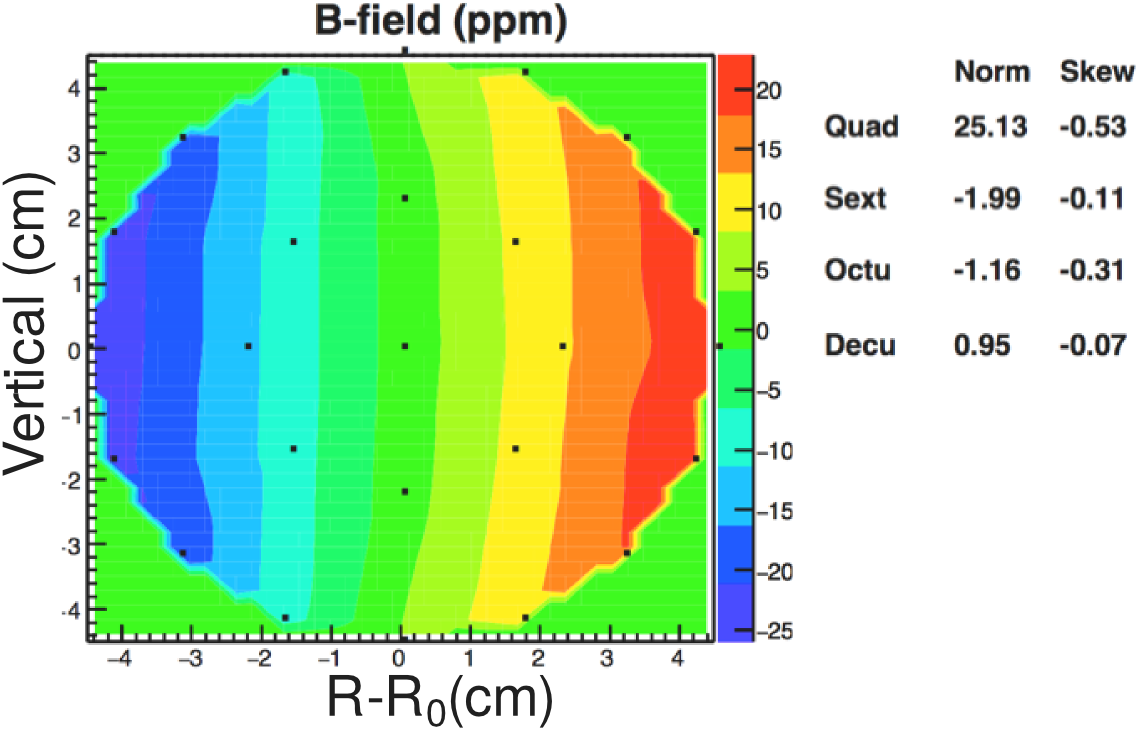}}\label{fig:b2}}
 \subfigure[]{\raisebox{0.\height}{\includegraphics[width=0.485\textwidth]{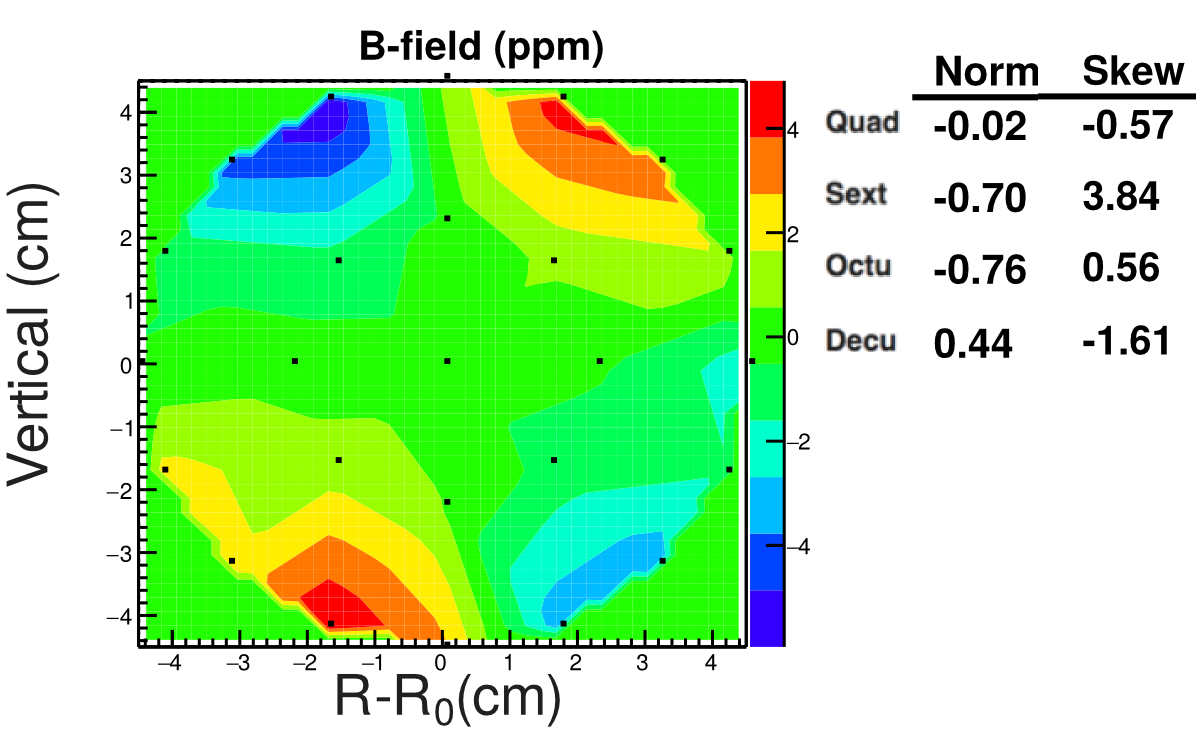}}\label{fig:b3}} 
\caption{\ref{fig:b1} Relative difference to the average magnetic field value as a function of the azimuth when the magnet was first powered-on in October 2015 (red),
in August 2016 (blue) and the goal (pink). Azimuthally averaged field as a function of the transverse coordinates in October 2015 \ref{fig:b2} and August 2016 \ref{fig:b3}.}
\label{fig:b}
\end{figure}

Given the required sensitivity, pulsed nuclear magnetic resistance (NMR) techniques are used. Tremendous progress were achieve between October 2015 and August 
2016 regarding the passive shimming\footnote{Passive shimming refers to the set of mechanical adjustments that are performed during the assembly of the ring and 
remain fixed during the running period.} of the magnet to produce as uniform field as possible (item 1). The first shimming step consisted in minimizing step and  
tilt discontinuity between adjacent pole pieces by moving the poles pieces adding metal shims under their mounting feet. The second step reduced the quadrupole 
component of the field. 48 iron top hats which affect the field in a 30 degree section and 864 wedges which affect a 10 degree section were moved through many iteration 
to reduce field variation around the ring. The last step is consisting in deploying 8,000 thin iron foils which affect 1 degree section of the ring on the pole surface 
(lower pole surface for the upper pole and vice-versa). Field simulations along with iterating the procedure will allow to reach the desired precision. Figure~\ref{fig:b} shows 
the NMR field 
scanning performed for the first time in October 2015 along with the most recent scan in August 2016. Figure~\ref{fig:b1} shows the relative difference to the average magnetic 
field value as a function of the azimuth. The relative difference dropped from $\pm$700 ppm to $\pm$25 ppm with the nominal goal being $\pm$25 ppm. Figures~\ref{fig:b2} 
and~\ref{fig:b3}  show the azimuthally averaged as a function of the transverse coordinates for respectively October 2015 and August 2016. We can see that the $\pm$25 ppm variation 
with its dipole contribute dropped 
to $\pm$3 ppm with the main contribution being now the sextupole and octupole components of the field. The nominal goal is $\pm$1 ppm. 

\section{Conclusion}

The Fermilab E989 experiment is on schedule to release its first result (with the BNL E821 statistics) of the measurement of the anomalous magnetic moment of the 
muon in Spring 2018. The modification of Fermilab accelerator complex will be done by Spring 2017 when the commissioning of the storage ring will start. By that 
time the various detector subsystems will be installed with the ring inside the experimental hall. The passive shimming of the magnet to produce a highly uniform 
magnetic field is currently well underway and should complete in the following months, leaving the floor open for the active shimming to start. The final result anticipated for 
2020 aims to reduce the total uncertainty of the BNL E821 experiment by a factor four. 

\section{Acknowledgement}

This research was supported, in part, by the U.S. National Science Founda-
tion’s MRI program PHY-1337542 and by the U.S. Department of Energy Office of Science, Office
of Nuclear Physics under award number DE-FG02-97ER41020.

\end{document}